\documentclass[aip,jcp,preprint,amsmath]{revtex4-1}
\usepackage{graphicx,dcolumn,bm,color}

\newcommand{\br}{\mathbf{r}}
\newcommand{\bu}{\mathbf{u}}
\newcommand{\bo}{\mathbf{0}}
\newcommand{\Ec}{\varepsilon_c}
\newcommand{\EHF}{E_\text{HF}}
\newcommand{\EL}{E_c^\text{LDA1}}
\newcommand{\EgL}{E_c^\text{gLDA1}}
\newcommand{\EP}{E_c^\text{MP2}}
\newcommand{\EPP}{E_c^\text{MP3}}
\newcommand{\ECI}{E_c^\text{FCI}}

\newcommand{\G}{\Gamma}
\newcommand{\mc}{\multicolumn}
\newcommand{\mD}{\mathcal{D}}
\newcommand{\Eh}{E_\text{h}}
\newcommand{\mEh}{\text{m}E_\text{h}}
\newcommand{\uEh}{\mu E_\text{h}}
\newcommand{\nC}{_n\mathrm{C}}
\newcommand{\rs}{r_s}
\newcommand{\sr}{\sum r_{ij}}
\newcommand{\alert}[1]{\textcolor{black}{#1}}

\begin{document}

\title[Generalized Local Density Approximation] {Uniform Electron Gases. \\ II.  The Generalized Local Density Approximation in One Dimension}

\author{Pierre-Fran{\c c}ois Loos}
\email{pf.loos@anu.edu.au}
\author{Caleb J. Ball}
\author{Peter M. W. Gill}
\email{peter.gill@anu.edu.au}
\affiliation{Research School of Chemistry, Australian National University, Canberra ACT 0200, Australia}

\begin{abstract}
We introduce a generalization (gLDA) of the traditional Local Density Approximation (LDA) within density functional theory.  The gLDA uses both the one-electron Seitz radius $\rs$ and a two-electron hole curvature parameter $\eta$ at each point in space.  The gLDA reduces to the LDA when applied to the infinite homogeneous electron gas but, unlike the LDA, is is also exact for finite uniform electron gases on spheres.  We present an explicit gLDA functional for the correlation energy of electrons that are confined to a one-dimensional space and compare its accuracy with LDA, second- and third-order M{\o}ller-Plesset perturbation energies and exact calculations for a variety of inhomogeneous systems.
\end{abstract}


\maketitle

\section{Local Density Approximation}
The local density approximation (LDA), unlike most of the ``sophisticated'' density functional approximations in widespread use today, is truly a first-principles quantum mechanical method.\cite{ParrBook}  It is entirely non-empirical, depending instead on the properties of one of the great paradigms of modern physics:  the infinite homogeneous electron gas (HEG).\cite{Fermi26, Thomas27}  Application of the LDA is straightforward, at least in principle.  Although the electronic charge density $\rho(\br)$ in any real system is non-uniform, the LDA proceeds by assuming that the charge in an infinitesimal volume element around the point $\br$ behaves like a locally homogeneous gas of density $\rho(\br)$, and adds all of the resulting contributions together.  This implicitly assumes that the infinitesimal contributions are independent (which is undoubtedly not the case) but then requires only that the properties of the HEG be known for all values of $\rho$.

The density of a HEG is commonly given by $\rho$ (the number of electrons per unit volume) or the Seitz radius $\rs$ and these equivalent parameters are related by
\begin{equation}	\label{eq:rsrho}
	\rs^\mD \rho = \pi^{-\mD/2} \G(\mD/2+1)
\end{equation}
where $\mD$ is the dimensionality of the space in which the electrons move.  In terms of these, the LDA correlation functional is
\begin{equation}	\label{eq:EcLDA}
	E_c^{\rm LDA} = \int \rho(\br) \Ec(\rs(\br)) \,d\br
\end{equation}
where the correlation kernel $\Ec(\rs)$ is the reduced (\emph{i.e.}~per electron) correlation energy of the HEG with Seitz radius $\rs$.

In high-density HEGs ({\it i.e.}~$\rs \ll 1$), the kinetic energy dominates the Hamiltonian and the Coulomb repulsion between the electrons can be treated via perturbation theory.  This has facilitated investigations of $\Ec(\rs)$ in 3D\cite{Wigner34, Macke50, Bohm53, Pines53, GellMann57, DuBois59, Carr64, Misawa65, Onsager66, Isihara75, Isihara76, Wang91, Hoffman92, Endo99, Ziesche05, 3DEG11} and 2D\cite{Zia73, Glasser77, Rajagopal77, Isihara77, Isihara78, Isihara80, Glasser84, Seidl04, Giuliani07, 2DEG11} but, because the Coulomb operator is so strong in 1D that two electrons cannot touch, the 1D gas has received less attention.\cite{Fogler05, Astrakharchik11, UEG1D}

In low-density HEGs ({\it i.e.}~$\rs \gg 1$), the potential energy dominates, the electrons localize into a Wigner crystal and strong-coupling methods can be used to find asymptotic expansions of $\Ec(\rs)$.  Here, too, the 3D,\cite{Maradudin60, Carr61, Fein61} 2D\cite{Meissner76, Maradudin77} and 1D\cite{Fogler05} HEGs have all been studied.

For intermediate densities, the best estimates of $\Ec(\rs)$ come from Quantum Monte Carlo (QMC) calculations, as pioneered by Ceperley and refined by several other groups.\cite{Ceperley78, Ceperley80, Ceperley89, Kwon93, Ortiz94, Senatore96, Kwon98, Ortiz99, Varsano01, Foulkes01, Attaccalite02, Ceperley02, Mitas06, Drummond09, Luchow10, Alavi12}  By combining these with the high- and low-density results, various groups\cite{VWN80, Perdew81, Perdew92, Sun10} have constructed interpolating functions that allow $\Ec(\rs)$ to be estimated rapidly for any value of $\rs$.

Unfortunately, this approach is flawed, for the correlation energy of a uniform electron gas depends on more than just its $\rs$ value.\cite{UEGs12} We have therefore argued that $\Ec(\rs)$ should be generalized to $\Ec(\rs,\eta)$, where the parameter $\eta$ measures the \emph{two-electron} density.  Although not mathematically mandated,\cite{PerezJorda01} we prefer that $\eta$, like $\rs$, be a \emph{local} quantity.  In Section \ref{sec:eta}, we propose a definition for $\eta$ inspired by a number of previous researchers.\cite{Colle75, Stoll80, Becke83, Luken84, Dobson91}

To learn more about the two-parameter kernel, we have turned to the finite uniform electron gases (UEGs) formed when $n$ electrons are confined to a $\mD$-sphere.\cite{TEOAS09, QuasiExact09, Frontiers10, ExSpherium10, EcProof10, Glomium11, QR12, Ringium13}  In Section \ref{sec:ringium}, we report accurate values of $\eta$ and $\Ec(\rs,\eta)$ for electrons on a 1-sphere, systems that we call ``$n$-ringium''.  In Section \ref{sec:GLDA1}, we devise three functionals to approximate these results and in Section \ref{sec:valid}, we test two of these on small 1D systems.  Atomic units are used throughout.

\section{Hole curvature} \label{sec:eta}
Suppose that an electron lies at a point $\br$.  The probability $P(\bu|\br)$ that a second electron lies at $\br+\bu$ is given\cite{Coulson61, Coleman67, Colle75, Stoll80, Perdew92a, Cioslowski98, Pu99, Insights00, Overview03, Omega06, AnnuRep11, Proud13} by the conditional intracule
\begin{equation}
	P(\bu|\br) = \rho_2(\br,\br+\mathbf{u}) / \rho(\br) \alert{\ = \left[ \rho(\br+\mathbf{u}) + \rho_\text{xc}(\br,\br+\mathbf{u}) \right] / 2}
\end{equation}
where \alert{$\rho_\text{xc}$ is the exchange-correlation hole\cite{ParrBook} and}
\begin{equation}
	\rho_2(\br_1,\br_2) = n(n-1) \int |\Psi|^2 \ ds_1 ds_2 d\br_3 \ldots d\br_n
\end{equation}
is the spinless second-order density matrix.\cite{DavidsonBook}  For fixed $\br$, we have the normalization
\begin{equation}
	\int P(\bu|\br) \,d\bu = n-1
\end{equation}
Because the Laplacian $\nabla^2_\bu P(\bo|\br)$ measures the tightness of the hole around the electron at $\br$ and has dimensions of $1/(\text{Length})^{\mD+2}$, we can use the dimensionless hole curvature
\begin{equation} \label{eq:eta}
	\eta(\br) = C_\mD \,\rs(\br)^{\mD+2} \,\nabla^2_\bu P(\bo|\br)
\end{equation}
to measure the proximity of other electrons to one at $\br$.  (We will fix the coefficient $C_\mD$ in the next Section.)  It is difficult to find this Laplacian for the exact wave function but, at the Hartree-Fock (HF) level, it involves simple sums over the occupied orbitals, \textit{viz.}
\begin{equation} \label{eq:nablaHF}
	\nabla^2_\bu P(\bo|\br) = 2 \sum_i^\text{occ} | \nabla\psi_i |^2 - \frac{|\nabla\rho|^2}{2\rho}
\end{equation}
and we will therefore employ HF curvatures henceforth.\footnote{\alert{In response to a referee, we have calculated the exact and HF curvatures in 2-ringium for several values of the ring radius $R$.  The HF curvature is $\eta_\text{HF} = 3/4$ for all $R$.  In contrast, the exact curvature diminishes from $\eta = 3/4$ at $R = 0$, to $\eta = 9\pi/(32+12\pi) \approx 0.406$ at $R = 1/2$, to $\eta = 18\pi/(64\sqrt{6}+51\pi) \approx 0.178$ at $R = \sqrt{3/2}$, and finally to $\eta = 0$ at $R = \infty$.}}  The interesting connection between the curvature and the kinetic energy density \cite{Becke83} is worth noting.

\section{Calculations on $n$-Ringium} \label{sec:ringium}
\subsection{Density and curvature}
The HF orbitals of the ground state of $n$ electrons on a ring of radius $R$ are complex exponentials\cite{QR12, Ringium13}
\begin{gather}
	\psi_m(\theta) = (2\pi R)^{-1/2} \exp(im\theta)	\\
	m = -\frac{n-1}{2}, -\frac{n-3}{2}, \ldots , +\frac{n-3}{2}, +\frac{n-1}{2}
\end{gather}
and, because of the symmetry of the system, the density and Seitz radius
\begin{gather}
	\rho = n / (2\pi R)	\\
	\rs = \pi R / n
\end{gather}
do not depend on $\theta$.  The hole curvature is also constant and, using \eqref{eq:eta} and \eqref{eq:nablaHF}, one finds
\begin{align}
	\eta	& = 2C_1 (\pi R / n)^3 \sum_m^\text{occ} m^2 / (2\pi R^3)	\notag	\\
		& = C_1 (\pi^2/12) (1 - 1/n^2)
\end{align}
If we choose $C_1 = 12/\pi^2$ so that $\eta = 1$ for the 1D HEG (\emph{i.e.}~$\infty$-ringium), we obtain
\begin{equation} \label{eq:etaringium}
	\eta = 1 - 1/n^2
\end{equation}
In general, requiring that $\eta = 1$ in the $\mD$-dimensional HEG leads (via Fermi integration) to
\begin{equation} \label{eq:CD}
	C_\mD = \frac{(1+2/\mD)\pi^{\mD/2}/8}{\Gamma(1+\mD/2)^{1+4/\mD}}
\end{equation}
and the particular values $C_2 = \pi/4$ and $C_3 = (10/27)(4\pi/3)^{1/3}$.

\subsection{Correlation energy}
The Hamiltonian for $n$ electrons on a ring is
\begin{equation}
	\hat{H} = \hat{T} + \hat{V} = - \frac{1}{2} \sum_{i=1}^n \nabla_i^2 + \sum_{i<j}^n r_{ij}^{-1}
\end{equation}
where $r_{ij}$ is the distance (across the ring) between electrons $i$ and $j$.  As noted previously,\cite{Ringium13} the energy is independent of the spin-state and so we assume that all electrons are spin-up.  The exact wave function can then be written as $\Psi = F\Phi$, where the correlation factor
\begin{equation}
	F = \sum_{a=1}^\infty x_a f_a
\end{equation}
is a sum of functions $f_a$ which are $m_a$-term symmetric polynomials in the $r_{ij}$ (see Table \ref{tab:fi}) and $\Phi$ is the HF wave function\cite{Ringium13}
\begin{equation}
	\Phi = \frac{1}{\sqrt{n! (2\pi)^n}} \prod_{i<j}^n \hat{r}_{ij}
\end{equation}

\begin{table}
\caption{Definitions $f_a$ and number of terms $m_a$ in the correlation factors of degree 0, 1, 2, 3} \label{tab:fi}
\begin{ruledtabular}
\begin{tabular}{ccccccccc}
	& \mc{2}{c}{Degree 0}	&	\mc{2}{c}{Degree 1}	&		\mc{2}{c}{Degree 2}	&				\mc{2}{c}{Degree 3}		\\
		\cline{2-3}					\cline{4-5}						\cline{6-7}								\cline{8-9}
	&$f_1$	&	$m_1$	&	$f_2$	&	$m_2$		&$f_3 \dots f_5$&$m_3\ldots m_5$&$f_6 \ldots f_{13}$ & $m_6\ldots m_{13}$	\\
		\hline
	&$1$	&	1		&	$\sr$	&	$\nC_2$	&	$\sr^2$		&	$\nC_2$		&	$\sr^3$				&	$\nC_2$		\\
	&		&			&			&				&	$\sr r_{ik}$	&	$6\ \nC_3$		&	$\sr r_{ik} r_{jk}$	&	$6\ \nC_3$		\\
	&		&			&			&				&	$\sr r_{kl}$	&	$6\ \nC_4$		&	$\sr^2 r_{ik}$		&	$18\ \nC_3$	\\
	&		&			&			&				&				&					&	$\sr^2 r_{kl}$		&	$18\ \nC_4$	\\
	&		&			&			&				&				&					&	$\sr r_{ik} r_{il}$	&	$24\ \nC_4$	\\
	&		&			&			&				&				&					&	$\sr r_{ik} r_{jl}$	&	$72\ \nC_4$	\\
	&		&			&			&				&				&					&	$\sr r_{ik} r_{lm}$	&	$180\ \nC_5$	\\
	&		&			&			&				&				&					&	$\sr r_{kl} r_{mn}$	&	$90\ \nC_6$	\\
		\hline
Total&	1	&	1		&	$\sr$	&	$\nC_2$	&	$(\sr)^2$	&	$(\nC_2)^2$	&	$(\sr)^3$			&	$(\nC_2)^3$	\\
\end{tabular}
\end{ruledtabular}
\end{table}

Judicious integration by parts allows us to partition the total energy
\begin{equation}
	E = \frac{\langle \Psi | \hat{H} | \Psi \rangle}{\langle \Psi | \Psi \rangle}
\end{equation}
into the HF energy\cite{Ringium13}
\begin{equation}
	\EHF = T_\text{HF} + V_\text{HF} = \frac{n(n^2-1)}{24R^2} + \frac{1}{4\pi R} \left( \sum_{k=1}^n \frac{4n^2-1}{2k-1} - 3n^2 \right)
\end{equation}
and the correlation energy
\begin{equation} \label{eq:Ec}
	E_c = \frac{\langle \Phi \ | \ \frac{1}{2} \nabla F \cdot \nabla F + (\hat{V} - V_\text{HF})F^2 \ | \ \Phi \rangle}{\langle \Phi | F^2 | \Phi \rangle}
\end{equation}
$E_c$ can be minimized either by QMC methods\cite{Luchow10} or via the secular equation
\begin{equation}
	(\mathbf{T}+\mathbf{V}) \mathbf{x} = E_c\,\mathbf{S}\, \mathbf{x}
\end{equation}
where the overlap, kinetic and Coulomb matrix elements
\begin{subequations} \label{eq:STV}
	\begin{gather}
		S_{ab} = \langle \Phi | f_a  f_b | \Phi \rangle									\\
		T_{ab} = \frac{1}{2} \langle \Phi |  \nabla f_a \cdot \nabla f_b | \Phi \rangle	\\
		V_{ab} = \langle \Phi | f_a \hat{V} f_b | \Phi \rangle - V_\text{HF} S_{ab}
	\end{gather}
\end{subequations}
can be found analytically in Fourier space (Appendix A).  \alert{We have used the CASINO QMC package\cite{CASINO10} and, where possible, the Knowles--Handy Full CI program to confirm results.\cite{Knowles84, Knowles89}}

Table \ref{tab:Ec} shows the resulting near-exact correlation energies for ground-state $n$-ringium.  \alert{(Where these energies differ from those in Table VI of Ref.~\onlinecite{Ringium13}, the new values are superior.)}  The fact that the $\Ec$ values in a given column are not equal demonstrates that the correlation energy of a UEG is not determined by its $\rs$ value alone.\cite{UEGs12}  Moreover, the variations in $\Ec$ for a given $\rs$ are large:  the $n=2$ values, for example, are only about half of the $n = \infty$ values, implying that the correlation energy of a few-electron system is grossly overestimated by the LDA functional which is based on the HEG.

\begin{table}
\caption{$\eta$ and $-\Ec(\rs,\eta)$ ($\mEh$ per electron) for the ground state of $n$ electrons on a ring} \label{tab:Ec}
\footnotesize
\begin{ruledtabular}
\begin{tabular}{ccccccccccccc}
			&			&														\mc{11}{c}{$\rs$}															\\
																					\cline{3-13}
	$n$		&	$\eta$	&	0		&	1/10	&	1/5		&	1/2		&	1		&	2		&	5		&	10		&	20		&	50		&	100		\\
	\hline
	1		&	0		&	0		&	0		&	0		&	0		&	0		&	0		&	0		&	0		&	0		&	0		&	0		\\
	2		&	3/4		&	13.212	&	12.985	&	12.766	&	12.152	&	11.250	&   \ \ 9.802	&   \ \ 7.111	&	4.938	&	3.122	&	1.533	&	0.848	\\
	3		&	8/9		&	18.484	&	18.107	&	17.747	&	16.755	&	15.346	&	13.179	&   \ \ 9.369 &	6.427	&	4.030	&	1.965	&	1.083	\\
	4		&	15/16	&	21.174	&	20.700	&	20.250	&	19.027	&	17.324	&	14.765	&	10.391	&	7.087	&	4.425	&	2.150	&	1.184	\\
	5		&	24/25	&	22.756	&	22.216	&	21.706	&	20.332	&	18.444	&	15.648	&	10.947	&	7.441	&	4.636	&	2.249	&	1.237	\\
	6		&	35/36	&	23.775	&	23.190	&	22.638	&	21.161	&	19.148	&	16.196	&	11.285	&	7.655	&	4.774	&	2.307	&	1.268	\\
	7		&	48/49	&	24.476	&	23.855	&	23.273	&	21.723	&	19.618	&	16.557	&	11.509	&	7.795	&	4.844	&	2.345	&	1.289	\\
	8		&	63/64	&	24.981	&	24.328	&	23.729	&	22.122	&	19.951	&	16.813	&	11.664	&	7.890	&	4.901	&	2.370	&	1.302	\\
	9		&	80/81	&	25.360	&	24.686	&	24.067	&	22.415	&	20.199	&	17.001	&	11.777	&	7.960	&	4.941	&	2.389	&	1.312	\\
	10		&	99/100	&	25.651	&	24.960	&	24.327	&	22.644	&	20.386	&	17.143	&	11.857	&	8.013	&	4.973	&	2.404	&	1.320	\\
	$\infty$	&	1		&	27.416	&	26.597	&	25.91\ \ &	23.962	&	21.444	&	17.922	&	12.318	&	8.292	&	5.133	&	2.476	&	1.358	\\
\end{tabular}
\end{ruledtabular}
\end{table}

\section{Generalized Local Density Approximation} \label{sec:GLDA1}
In the LDA, the correlation contribution is estimated from $\rs$ alone, according to Eq.~\eqref{eq:EcLDA}.  However, the fact that UEGs with the same $\rs$, but different $\eta$, have different energies compels us to devise a Generalized Local Density Approximation (GLDA) wherein we write
\begin{equation}	\label{eq:EcGLDA}
	E_c^{\rm GLDA} = \int \rho(\br) \,\Ec(\rs(\br),\eta(\br)) \,d\br
\end{equation}
where the correlation kernel $\Ec(\rs,\eta)$ is the reduced correlation energy of a UEG with Seitz radius $\rs$ and curvature $\eta$.  For present purposes, we will use $\rs$ and $\eta$ values from the HF, rather than the exact, wave function.

One might think that the kernel could be constructed by fitting the results in Table \ref{tab:Ec} but these data allow us to construct $\Ec(\rs,\eta)$ only for $\eta \le 1$.  To construct the rest of the kernel will require accurate correlation energies for uniform gases with high curvatures ($\eta > 1$) but, \alert{although these arise in \emph{excited} states of $n$-ringium, this raises some fundamental questions which lie outside the scope of the present manuscript and will be discussed elsewhere.}

\subsection{High densities}
Rayleigh-Schr\"odinger perturbation theory for $n$-ringium yields the high-density expansion
\begin{align} \label{eq:highexp}
	\Ec(\rs,n) = \alpha_2(n) + \alpha_3(n) \rs + \alpha_4(n) \rs^2 + \ldots,									&&		(\rs \ll 1)
\end{align}
The leading coefficient\cite{Ringium13} is
\begin{align}
	\alpha_2(n)	& = -\frac{1}{n} \sum_{a<b}^\text{occ} \sum_{r=r_\text{min}}^\infty \frac{V_{r-a,r-b}^2}{(r-a)(r-b)}		\notag	\\
				& = - \frac{\pi^2}{360} + \frac{a\ln^2 n+b\ln n+c}{n^2} + \ldots
\end{align}
but, \alert{if we fit a truncated version of this series, while ensuring that $\alpha_2$ vanishes for one electron}, we obtain the approximation
\begin{align}
	\tilde\alpha_2(n) = - \frac{\pi^2}{360} \left(1-\frac{1}{n^2}\right) + \frac{\ln^2 n+3\ln n}{87n^2}			&&		(1 \le n < \infty)
\end{align}
which can be rewritten in terms of the curvature, using Eq.~\eqref{eq:etaringium} to obtain 
\begin{align} \label{eq:alpha}
	\tilde\alpha_2(\eta) = - \frac{\pi^2}{360} \eta + (1-\eta) \frac{\ln^2(1-\eta) - 6\ln(1-\eta)}{348}				&&		(0 \le \eta \le 1)
\end{align}
The accuracy of this approximation is shown in columns 2 and 3 of Table \ref{tab:maxerror}.

\subsection{Low densities}
Strong-coupling perturbation theory for $n$-ringium yields the low-density expansion
\begin{align} \label{eq:lowexp}
	\Ec(\rs,n) = \frac{\beta_2(n)}{\rs} + \frac{\beta_3(n)}{\rs^{3/2}} + \frac{\beta_4(n)}{\rs^2} + \ldots			&&		(\rs \gg 1)
\end{align}
The leading coefficient is the difference between the Wigner crystal Coulomb coefficient
\begin{align}
	E_\text{V}^\text{W}(n)	& = \frac{\pi}{4n} \sum_{k=1}^{n-1} \csc(k\pi/n)								\notag	\\
							& = \frac{1}{2} \int_0^1 \frac{1-x^{n-1}}{1-x} \frac{dx}{1+x^n}				\notag	\\
							& = \frac{\ln n}{2} + \frac{\gamma+\ln(2/\pi)}{2} - \frac{\pi^2}{144n^2} + \ldots
\end{align}
and the HF Coulomb coefficient
\begin{align}
	E_\text{V}^\text{HF}(n)	& = \left(1-\frac{1}{4n^2}\right) \sum_{k=1}^n \frac{1}{2k-1} - \frac{3}{4}		\notag	\\
							& = \left(1-\frac{1}{4n^2}\right) \left(\frac{\ln n}{2} + \frac{\gamma+2\ln2}{2} + \frac{1}{48n^2} + \ldots\right) - \frac{3}{4}
\end{align}
It follows that
\begin{equation}
	\beta_2(n) = \frac{3}{4} - \frac{\ln 2\pi}{2} + \frac{\ln n}{8n^2} + \frac{18\gamma+36\ln2-3-\pi^2}{144n^2} + \ldots
\end{equation}
but, if we truncate this series after the $n^{-2}$ term and modify it to ensure that $\beta_2$ vanishes for one electron, we obtain the approximation
\begin{align}
	\tilde\beta_2(n) = \left(\frac{3}{4} - \frac{\ln 2\pi}{2}\right)\left(1-\frac{1}{n^2}\right) + \frac{\ln n}{8n^2}	&&		(1 \le n < \infty)
\end{align}
which can be rewritten in terms of the curvature, using Eq.~\eqref{eq:etaringium} to obtain 
\begin{align} \label{eq:beta}
	\tilde\beta_2(\eta) = \left(\frac{3}{4} - \frac{\ln 2\pi}{2}\right) \eta - \frac{(1-\eta) \ln(1-\eta)}{16}			&&		(0 \le \eta \le 1)
\end{align}
The accuracy of this approximation is shown in columns 4 and 5 of Table \ref{tab:maxerror}.

\subsection{Intermediate densities}
How can we model $\Ec(\rs,\eta)$ for fixed $\eta$?  Ideally, we would like a function that reproduces the behaviors of Eqs~\eqref{eq:highexp} and \eqref{eq:lowexp} and interpolates accurately between these limits.  However, for practical reasons, we will content ourselves with a function that approaches $\tilde\alpha_2(\eta)$ for small $\rs$, behaves like $\tilde\beta_2(\eta)/\rs$ for large $\rs$, and changes monotonically between these.

Although we could use robust interpolation,\cite{Cioslowski12} the hypergeometric function\cite{NISTbook}
\begin{align}
	f(r)	& = \alpha F\left(1 , \frac{3}{2} , \gamma , \frac{2\alpha(1-\gamma)}{\beta} r \right)	\\
		& \sim	\begin{cases}
						\alpha + O(r)			&	r \ll 1	\\
						\beta / r + O(r^{-3/2})	&	r \gg 1
					\end{cases}
\end{align}
possesses all of the desired features and we therefore adopt the approximate kernel
\begin{equation} \label{eq:Ectilde}
	\tilde\Ec(\rs,\eta) = \tilde\alpha_2(\eta) F\left(1 , \frac{3}{2} , \tilde\gamma(\eta) , \frac{2\tilde\alpha_2(\eta)(1-\tilde\gamma(\eta))}{\tilde\beta_2(\eta)} \rs \right)
\end{equation}
Table \ref{tab:maxerror}  shows that this kernel models the energies in Table \ref{tab:Ec} well if we choose
\begin{align}
	\tilde\gamma(n) = \frac{19}{16} \left(\frac{4n-3}{2n-1}\right)									&&		(1 \le n < \infty)
\end{align}
or, equivalently,
\begin{align} \label{eq:gamma}
	\tilde\gamma(\eta) = \frac{19}{16} \left(\frac{4-3\sqrt{1-\eta}}{2-\sqrt{1-\eta}}\right)			&&		(0 \le \eta \le 1)
\end{align}
reproduces the Table \ref{tab:Ec} data to within a relative error of 1\% and absolute error of 0.20 $\mEh$.

\begin{table}
	\caption{Application of the $\tilde\Ec(\rs,\eta)$ approximation to the data in Table \ref{tab:Ec}} \label{tab:maxerror}
	\begin{ruledtabular}
		\begin{tabular}{ccccccccc}
					&			&				&					&				&					&			&	\mc{2}{c}{Max errors}	\\
																															\cline{8-9}
			$n$		&	$\eta$	&  $-\alpha_2$	& $-\tilde\alpha_2$&	$-\beta_2$	& $-\tilde\beta_2$	&$\tilde\gamma$&	\%		& Abs ($\mEh$)\\
			\hline
			2		&	3/4		&	0.01321	&	0.01321		&	0.1073		&	0.1050			&	1.9792	&	1.0		&		0.10	\\
			3		&	8/9		&	0.01848	&	0.01862		&	0.1361		&	0.1349			&	2.1375	&	0.9		&		0.13	\\
			4		&	15/16	&	0.02117	&	0.02133		&	0.1483		&	0.1475			&	2.2054	&	0.8		&		0.16	\\
			5		&	24/25	&	0.02276	&	0.02291		&	0.1546		&	0.1541			&	2.2431	&	0.8		&		0.18	\\
			6		&	35/36	&	0.02378	&	0.02391		&	0.1584		&	0.1580			&	2.2670	&	0.7		&		0.14	\\
			7		&	48/49	&	0.02448	&	0.02460		&	0.1608		&	0.1605			&	2.2837	&	0.8		&		0.16	\\
			8		&	63/64	&	0.02498	&	0.02509		&	0.1624		&	0.1622			&	2.2958	&	0.7		&		0.16	\\
			9		&	80/81	&	0.02536	&	0.02546		&	0.1636		&	0.1635			&	2.3051	&	0.7		&		0.15	\\
			10		&	99/100	&	0.02565	&	0.02574		&	0.1645		&	0.1644			&	2.3125	&	0.8		&		0.20	\\
			$\infty$	&	1		&	0.02742	&	0.02742		&	0.1689		&	0.1689			&	2.3750	&	0.8		&		0.13	\\
		\end{tabular}
	\end{ruledtabular}
\end{table}

\subsection{The LDA1, GLDA1 and gLDA1 functionals}
We can now consider three approximate kernels for correlation in 1D systems.  The first is the LDA1 kernel, which is defined by
\begin{equation} \label{eq:kerLDA}
	\Ec^\text{LDA1}(\rs) = \alpha F\left(1 , \frac{3}{2} , \tilde\gamma , \frac{2\alpha(1-\tilde\gamma)}{\beta} \rs \right)
\end{equation}
where $\alpha = -\pi^2/360$, $\beta = 3/4 - (\ln 2\pi)/2$ and $\tilde\gamma = 19/8$.  This underpins the traditional LDA and, by construction, it is exact (within fitting errors) for the 1D HEG or, equivalently, for $\infty$-ringium.  It is independent of the hole curvature $\eta$.

The second is the GLDA1 kernel, which is defined by
\begin{equation}
	\Ec^\text{GLDA1}(\rs,\eta) = \tilde\alpha_2(\eta) F\left(1 , \frac{3}{2} , \tilde\gamma(\eta) , \frac{2\tilde\alpha_2(\eta)(1-\tilde\gamma(\eta))}{\tilde\beta_2(\eta)} \rs \right)
\end{equation}
where $\tilde\alpha_2(\eta)$, $\tilde\beta_2(\eta)$ and $\tilde\gamma(\eta)$ are defined in Eqs \eqref{eq:alpha}, \eqref{eq:beta} and \eqref{eq:gamma}.  Unfortunately, because of a lack of information about high-curvature UEGs, these three equations are not defined for $\eta > 1$ and thus, at this time, the GLDA1 is defined only for systems where $\eta \le 1$ at all points.  Completing the definition of the GLDA1 is an important topic for future work.

The third is the gLDA1 kernel, a partially corrected LDA, which is defined by
\begin{equation} \label{eq:kerGLDA}
	\Ec^\text{gLDA1}(\rs,\eta) =	\begin{cases}
									\Ec^\text{GLDA1}(\rs,\eta)	&	\eta < 1	\\
									\Ec^\text{LDA1}(\rs)			&	\eta \ge 1
								\end{cases}
\end{equation}
When applied to UEGs with $\eta \ge 1$, the gLDA1 and LDA1 kernels are, of course, identical.  However, when applied to gases with $\eta < 1$, they behave differently and, by construction, the gLDA1 kernel is exact (within fitting errors) for any $n$-ringium.

The gLDA1 kernel defaults back to the LDA1 kernel at points where $\eta > 1$ but we cannot predict \textit{a priori} whether this will cause it to under-estimate or to over-estimate the GLDA.  If the monotonic increase in the magnitude of the kernel between $\eta = 0$ and $\eta = 1$ continues beyond $\eta = 1$, then the gLDA1 kernel (which assumes that the kernel is constant beyond $\eta = 1$) will underestimate the GLDA1 kernel and consequently underestimate the magnitude of the correlation energies in systems with high-curvature regions.

Until the true kernel for $\eta > 1$ is known, we cannot draw any firm conclusions about the accuracy of GLDA1.  However, it is reasonable to conjecture that even the imperfect gLDA1 may be superior to LDA1 for density functional theory (DFT) calculations on inhomogeneous 1D systems and we now explore this through some preliminary validation studies.

\section{Validation} \label{sec:valid}
Having defined the gLDA1 functional, we turn now to its validation.  The functional is exact by construction for any $n$-ringium, so we require systems with non-uniform densities.  There is no standard set of 1D models with accurately known correlation energies, so it was necessary to devise our own and we chose the ground states of $n$ electrons in a 1D box of length $L = \pi$ (a family that we call the $n$-boxiums) and of $n$ electrons in a 1D harmonic well with force constant $k = 1$ (a family that we call the $n$-hookiums).  Whereas the HOMO--LUMO gap in $n$-boxium increases roughly linearly with $n$, that in $n$-hookium slowly decreases.  We therefore regard them as ``large-gap'' and ``small-gap'' systems, respectively.

Given that the fitting errors (Table \ref{tab:maxerror}) in the gLDA1 functional can be of the order of 0.1 $\mEh$, we aimed to obtain the energies of the $n$-boxium and $n$-hookium to within 0.1 $\mEh$ of their complete basis set (CBS) limits.  This is easily achieved for the HF, LDA1 and gLDA1 energies, because they converge exponentially\cite{Kutzelnigg94,GaussExp12,Kutzelnigg13} with the size $M$ of the one-electron basis, but it is less straightforward for traditional post-HF energies.

We analysed the convergence behavior (see Appendix B) of M{\o}ller-Plesset perturbation (MP2 and MP3) and full configuration interaction (FCI) energies in 2-ringium, 2-boxium and 2-hookium and our results are summarised in Table \ref{tab:convergence}.  From these, we devised appropriate extrapolation formulae and applied these to the energies obtained with our largest basis sets.  We also used QMC calculations\cite{CASINO10} to assess the accuracy of our extrapolated FCI energies.

Tables \ref{tab:convbox} and \ref{tab:convhook} show the energies obtained for 5-boxium and 5-hookium, respectively, as the basis set size increases from $M = 5$ to $M = 30$.  The three components of the third-order energy are separated because of their different convergence behaviors.  Table \ref{tab:apps} summarizes our best estimates of the HOMO--LUMO gaps, together with the HF, LDA1, gLDA1, MP2, MP3 and FCI energies, for $n$-boxium and $n$-hookium with $n$ = 2, 3, 4 or 5.

\begin{table}
	\caption{Basis set truncation errors $\Delta E_M$ for the energies in two-electron systems} \label{tab:convergence}
	\begin{ruledtabular}
		\begin{tabular}{cccc}
						&		MP2		&		MP3			&		FCI			\\
			\hline
			2-ringium	&	$O(M^{-3})$	&	$O(M^{-3})$		&	$O(M^{-3})$	\\
			2-boxium	&	$O(M^{-3})$	&	$O(M^{-3})$		&	$O(M^{-3})$	\\
			2-hookium	&	$O(M^{-3/2})$	&	$O(M^{-3/2}/\ln M)$	&	$O(M^{-3/2})$	\\
		\end{tabular}
	\end{ruledtabular}
\end{table}

\subsection{$n$-Boxium}
\begin{figure}
	\includegraphics[width=0.49\textwidth]{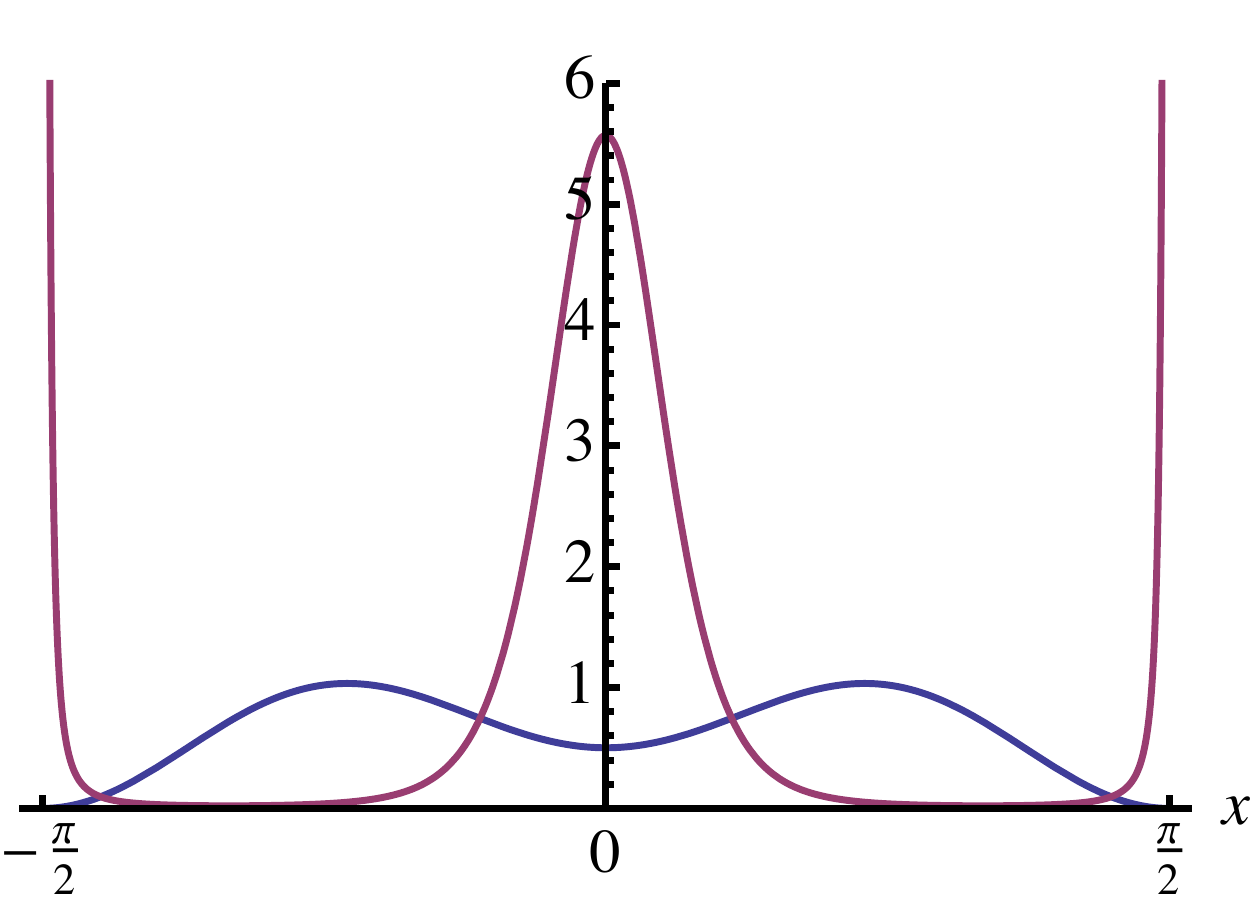}
	\includegraphics[width=0.49\textwidth]{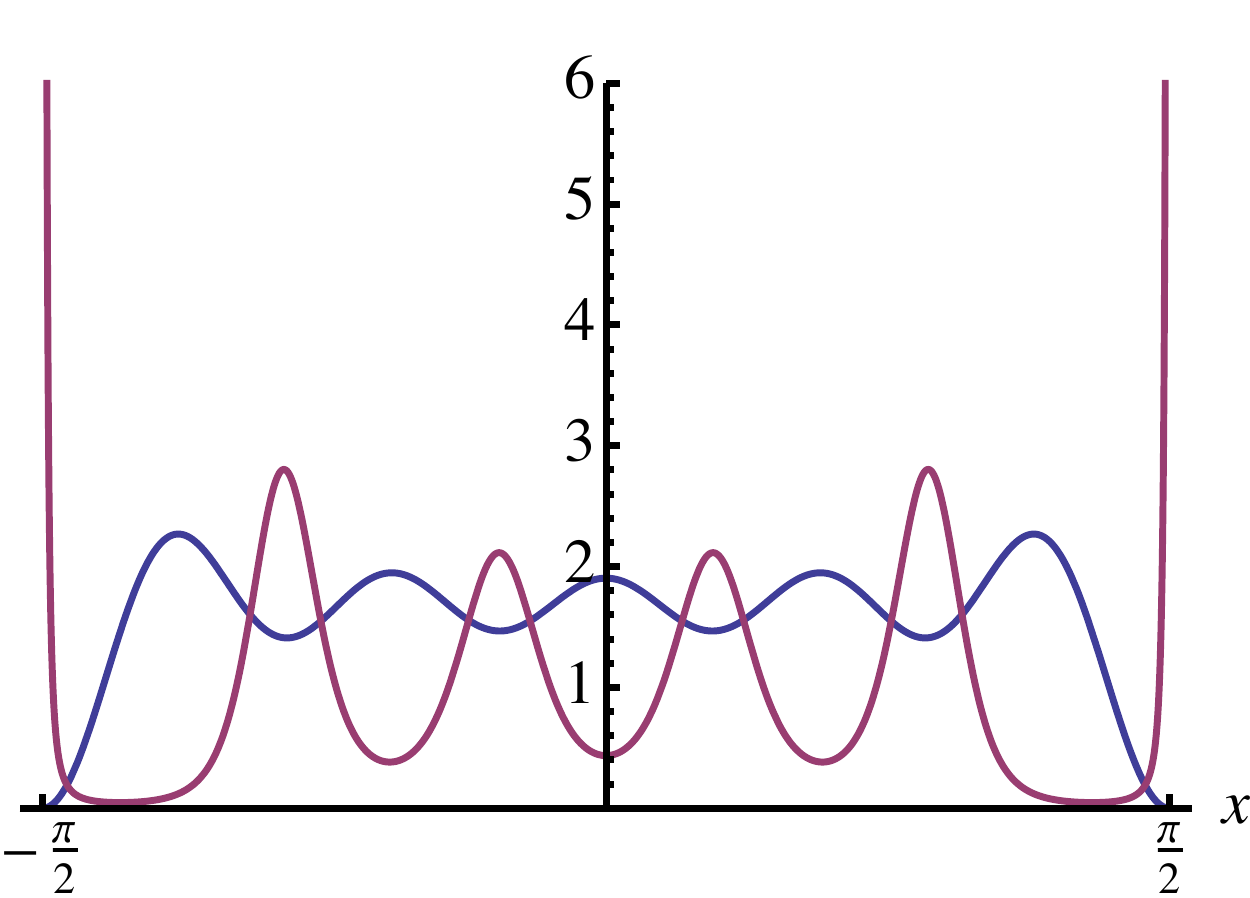}
	\caption{HF density $\rho(x)$ (blue) and curvature $\eta(x)$ (red) in 2-boxium (left) and 5-boxium (right)} \label{fig:boxium}
\end{figure}

The 2-boxium system (albeit with length $L = 3$) was studied in a basis of delta functions by Salter \textit{et al.}~\cite{Salter01} and, using 804609 basis functions,  they obtained energies within roughly 10 $\uEh$ of the exact values.  The present work is the first study of $n$-boxium with $n \ge 3$.

The orbitals of 1-boxium are
\begin{align}
	\phi_m(x) = \begin{cases}
					\sqrt{2/\pi} \cos m x		&	m\text{ is odd}	\\
					\sqrt{2/\pi} \sin m x		&	m\text{ is even}
				\end{cases}										&&		(m = 1,2,3,\ldots)
\end{align}
and the first $M$ of these form a convenient orthonormal basis for expanding the HF orbitals in $n$-boxium.  The antisymmetrized two-electron integrals $\langle \mu \sigma || \nu \lambda \rangle$ can be found in terms of the \alert{Sine and Cosine Integral} functions\cite{NISTbook} and we have used these to perform SCF calculations with up to $M = 30$ basis functions.  Our convergence criterion was $\max |[\mathbf{P},\mathbf{F}]| < 10^{-5}$.

We first discuss 2-boxium.  Choosing $M = 8$ yields the HF orbitals
\begin{subequations}
\begin{gather}
	\psi_1(x) = 0.994844 \,\phi_1(x) - 0.101256 \,\phi_3(x) - 0.005729 \,\phi_5(x) - 0.000044 \,\phi_7(x)	\\
	\psi_2(x) = 0.999715 \,\phi_2(x) - 0.023850 \,\phi_4(x) + 0.000728 \,\phi_6(x) - 0.000176 \,\phi_8(x)
\end{gather}
\end{subequations}
and Fig.~\ref{fig:boxium} reveals that the density $\rho$ has maxima at $x \approx \pm \pi/4$, indicating that an electron is likely to be found in these regions.  LDA1 interprets these maxima as the most strongly correlated regions in the well and, through Eqs \eqref{eq:EcLDA} and \eqref{eq:kerLDA}, predicts the correlation energy
\begin{equation}
	\EL = \int_{-\pi/2}^{\pi/2} \rho(x) \Ec^\text{LDA1}(\rs) \,dx = -46.1\ \mEh
\end{equation}
In contrast, because the hole curvature $\eta$ is strongly peaked at the center and edges of the box and is small near the density maxima, gLDA1 identifies the \emph{center} of the box as the most correlated region and Eqs \eqref{eq:EcGLDA} and \eqref{eq:kerGLDA} predict the much smaller correlation energy
\begin{equation}
	\EgL = \int_{-\pi/2}^{\pi/2} \rho(x) \Ec^\text{gLDA1}(\rs,\eta) \,dx = -11.0\ \mEh
\end{equation}
LDA1 and gLDA1 offer very different qualitative and quantitative descriptions of 2-boxium, but both perturbation theory ($\EP = -8.33\ \mEh$ and $\EPP = -9.45\ \mEh$) and near-exact calculations ($\ECI = -9.82\ \mEh$) support the gLDA1 picture.

We have also performed HF, LDA1, gLDA1, MP2, MP3 and FCI calculations on 3-, 4- and 5-boxium and the density and curvature for 5-boxium are shown on the right of Fig.~\ref{fig:boxium}.  Both functions oscillate much more rapidly but with much smaller amplitude than in 2-boxium, and it is easy to foresee that, as the number of electrons becomes large, both the density and the curvature will become increasingly uniform.

The convergence of the 5-boxium energies is shown in Table \ref{tab:convbox} and confirms the theoretical predictions of Table \ref{tab:convergence}.  The LDA1 energies, which depend only on the density $\rho(x)$, converge rapidly, changing by less than 1 $\uEh$ beyond $M = 11$.  The HF and gLDA energies, which depend on the orbitals (rather than the density) converge more slowly, achieving 1 $\uEh$ convergence around $M = 20$.  Because the occupied orbitals converge more rapidly than the virtual ones,\cite{DBMP11} the $O^4 V^2$ component of MP3 converges almost as fast as HF, the $O^3 V^3$ component (which is negative) converges more slowly, and the $O^2 V^4$ component (which is positive) even more slowly.\footnote{The symbols ``O'' and ``V'' refer to the number of occupied and virtual orbitals, respectively.  The $O^4 V^2$ component, for example, involves four sums over occupied orbitals and two over virtual orbitals.}  Because of the resulting differential cancellation,\cite{Noga94, Tew10} the total 3rd-order contribution initially becomes more negative, reaches a minimum at $M = 13$ and rises thereafter.  The MP2 energy is the most slowly converging, and changes by 60 $\uEh$ between $M = 29$ and $M = 30$.  It is interesting to note the almost perfectly linear growth of the third-order energies.  Because the $n$-boxiums are large-gap systems, MP2 and MP3 work well, recovering more than 92\% and 99\% of the correlation energy in 5-boxium.

Our best estimates of the CBS limit HF and correlation energies are summarized in the left half of Table \ref{tab:apps}.  Because LDA1 operates without the benefit of curvature information, it gravely overestimates the correlation energy, by between a factor of five (for 2-boxium) and a factor of just under two (for 5-boxium).  In contrast, gLDA1 is within 12\% of the true correlation energy for all $n$-boxiums studied.

\begin{table}
	\caption{Basis set convergence of $\EHF$ (in $\Eh$) and $E_c$ energies (in $\mEh$) in 5-boxium} \label{tab:convbox}
	\begin{ruledtabular}
		\begin{tabular}{ccccccccc}
					&					&				&				&				&		\mc{3}{c}{$E^{(3)}$ components}		&				\\
																											\cline{6-8}
			$M$	&	$\EHF$			&	$-\EL$		&	$-\EgL$	&	$-\EP$		&   $O^4 V^2$	&   $O^3 V^3$	&   $O^2 V^4$	&	$-\ECI$		\\
			\hline
			5		&	40.990\,531	&	126.517	&	68.858		&		0		&		0		&		0		&		0		&		0		\\
			6		&	40.855\,806	&	126.499	&	63.929		&		0		&		0		&		0		&		0		&		0		\\
			7		&	40.807\,556	&	126.486	&	63.678		&	16.020		&	1.129		&	$-$3.512	&	0.804		&	17.840		\\
			8		&	40.798\,066	&	126.482	&	63.314		&	28.753		&	1.728		&	$-$6.379	&	1.683		&	32.157		\\
			9		&	40.793\,901	&	126.478	&	63.208		&	38.619		&	2.085		&	$-$8.570	&	2.475		&	43.234		\\
			10		&	40.793\,518	&	126.478	&	63.207		&	45.564		&	2.276		&	$-$10.046	&	3.104		&	50.937		\\
			11		&	40.792\,520	&	126.477	&	63.067		&	49.972		&	2.371		&	$-$10.871	&	3.577		&	55.640		\\
			12		&	40.792\,237	&		"		&	63.024		&	53.055		&	2.426		&	$-$11.394	&	3.932		&	58.850		\\
			13		&	40.792\,064	&		"		&	63.026		&	55.272		&	2.458		&	$-$11.729	&	4.203		&	61.102		\\
			14		&	40.792\,057	&		"		&	63.031		&	56.876		&	2.478		&	$-$11.939	&	4.411		&	62.682		\\
			15		&	40.792\,051	&		"		&	63.019		&	58.059		&	2.491		&	$-$12.071	&	4.572		&	63.815		\\
			16		&	40.792\,051	&		"		&	63.017		&	58.946		&	2.499		&	$-$12.157	&	4.699		&	64.646		\\
			17		&	40.792\,049	&		"		&	63.026		&	59.624		&	2.505		&	$-$12.214	&	4.799		&	65.271		\\
			18		&	40.792\,049	&		"		&	63.027		&	60.151		&	2.509		&	$-$12.254	&	4.880		&	65.750		\\
			19		&	40.792\,049	&		"		&	63.028		&	60.568		&	2.512		&	$-$12.282	&	4.945		&	66.124		\\
			20		&	40.792\,048	&		"		&	63.028		&	60.901		&	2.514		&	$-$12.302	&	4.998		&	66.420		\\
			21		&		"			&		"		&	63.029		&	61.170		&	2.515		&	$-$12.317	&	5.042		&	66.658		\\
			22		&		"			&		"		&		"		&	61.391		&	2.516		&	$-$12.328	&	5.079		&	66.852		\\
			23		&		"			&		"		&		"		&	61.574		&	2.517		&	$-$12.337	&	5.110		&	67.011		\\
			24		&		"			&		"		&		"		&	61.726		&	2.517		&	$-$12.343	&	5.136		&	67.143		\\
			25		&		"			&		"		&		"		&	61.854		&	2.518		&	$-$12.349	&	5.158		&	67.253		\\
			26		&		"			&		"		&		"		&	61.963		&	2.518		&	$-$12.353	&	5.176		&	67.346		\\
			27		&		"			&		"		&		"		&	62.055		&	2.519		&	$-$12.356	&	5.193		&	67.425		\\
			28		&		"			&		"		&		"		&	62.134		&		"		&	$-$12.358	&	5.207		&	67.493		\\
			29		&		"			&		"		&		"		&	62.203		&		"		&	$-$12.360	&	5.219		&	67.551		\\
			30		&		"			&		"		&		"		&	62.262		&		"		&	$-$12.362	&	5.230		&	67.601		\\
		\end{tabular}
	\end{ruledtabular}
\end{table}

\subsection{$n$-Hookium}
\begin{figure}
	\includegraphics[width=0.49\textwidth]{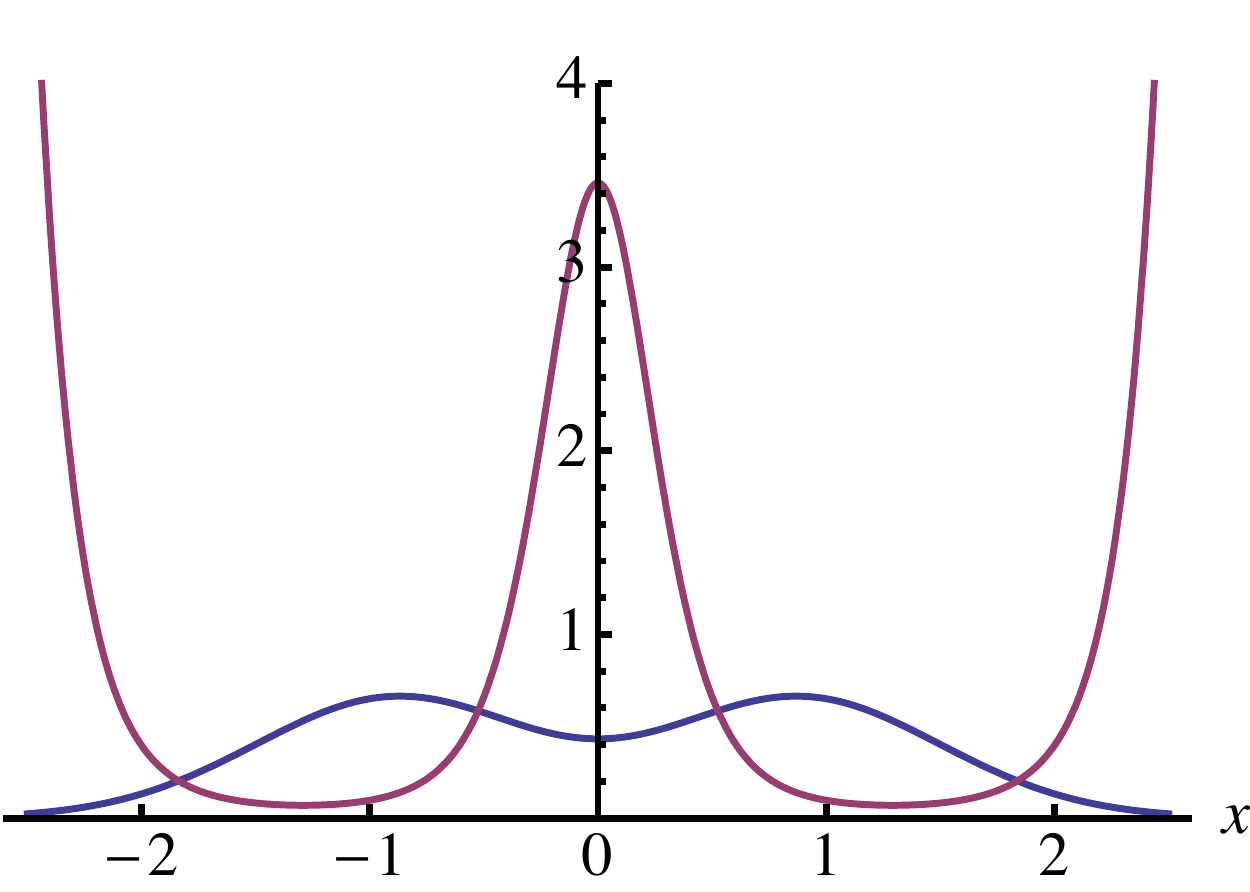}
	\includegraphics[width=0.49\textwidth]{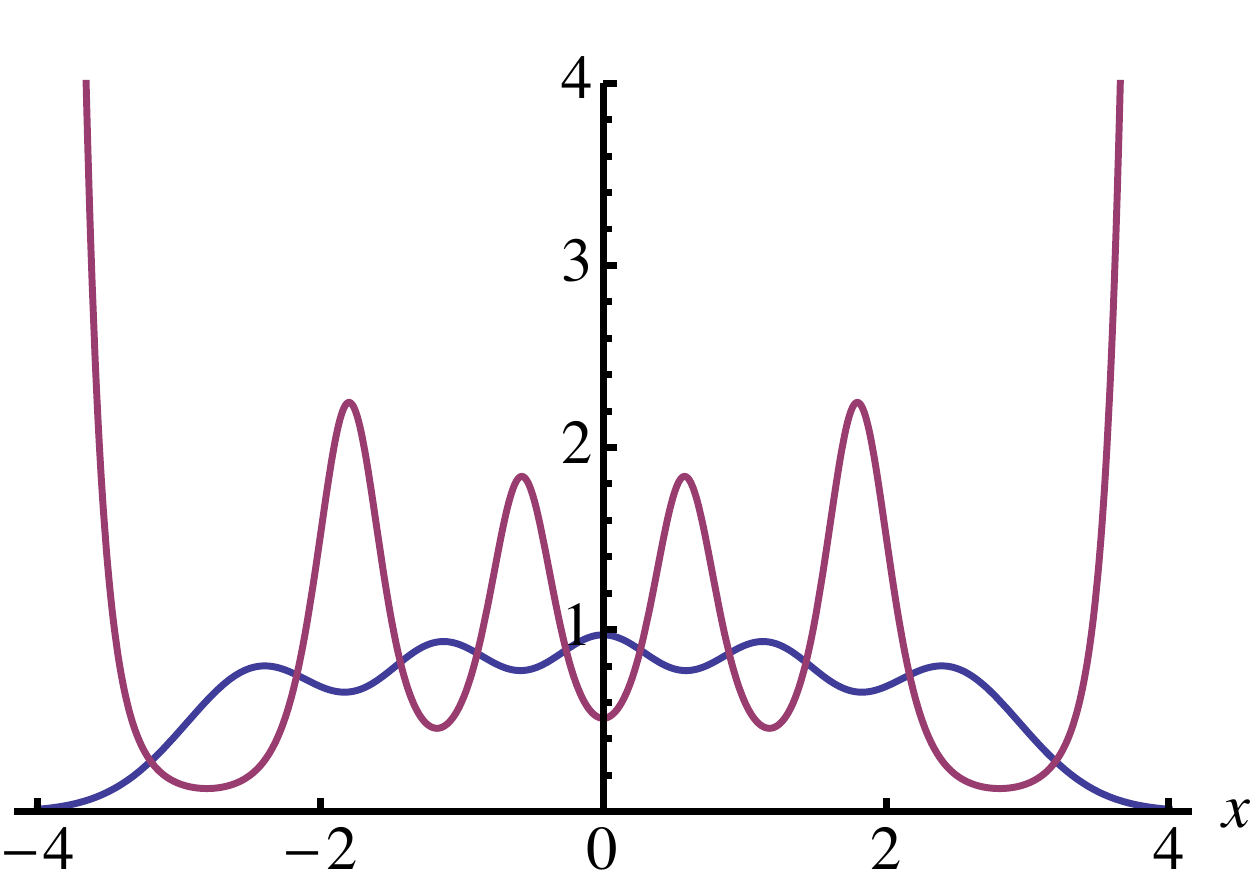}
	\caption{HF density $\rho(x)$ (blue) and curvature $\eta(x)$ (red) in 2-hookium (left) and 5-hookium (right)} \label{fig:hookium}
\end{figure}

Electrons in 3D harmonic wells have been studied by numerous authors\cite{Kestner62, Santos68, White70, Benson70, Kais89, Taut93, Ivanov99, Cioslowski00, Henderson01, Hookium03, Katriel05, HookCorr05, Ragot08, EcLimit09, Cioslowski13} but this is the first investigation of $n$ electrons in a 1D harmonic well.  The orbitals of 1-hookium are
\begin{align}
	\phi_m(x) = \frac{H_{m-1}(x) \exp(-x^2/2)}{\sqrt{\pi^{1/2}2^{m-1}(m-1)!}}		&&		(m = 1,2,3,\ldots)
\end{align}
and the first $M$ of these form a convenient orthonormal basis for expanding the HF orbitals in $n$-hookium.  The antisymmetrized two-electron integrals $\langle \mu \sigma || \nu \lambda \rangle$ can be found in closed form (\textit{e.g.}~see Appendix B) and we have used these to perform SCF calculations with up to $M = 30$ basis functions.  Our convergence criterion was $\max |[\mathbf{P},\mathbf{F}]| < 10^{-5}$.

\begin{table}
	\caption{Basis set convergence of $\EHF$ (in $\Eh$) and $E_c$ energies (in $\mEh$) in 5-hookium} \label{tab:convhook}
	\begin{ruledtabular}
		\begin{tabular}{ccccccccc}
					&					&				&				&				&		\mc{3}{c}{$E^{(3)}$ components}		&				\\
																											\cline{6-8}
			$M$	&	$\EHF$			&	$-\EL$		&	$-\EgL$	&	$-\EP$		&   $O^4 V^2$	&   $O^3 V^3$	&   $O^2 V^4$	&	$-\ECI$		\\
			\hline
			5		&	19.649\,014	&	116.419	&	75.381		&		0		&		0		&		0		&		0		&		0		\\
			6		&	19.353\,767	&	115.709	&	60.013		&		0		&		0		&		0		&		0		&		0		\\
			7		&	19.180\,033	&	114.892	&	64.207		&	18.983		&	2.783		&	$-$7.833	&	1.952		&	23.103		\\
			8		&	19.171\,222	&	114.736	&	63.602		&	27.047		&	3.466		&	$-$11.252	&	2.972		&	33.352		\\
			9		&	19.167\,260	&	114.619	&	63.990		&	33.786		&	4.058		&	$-$14.364	&	4.077		&	42.140		\\
			10		&	19.165\,782	&	114.658	&	63.679		&	37.870		&	4.298		&	$-$16.063	&	4.812		&	47.219		\\
			11		&	19.165\,244	&	114.680	&	63.512		&	41.400		&	4.488		&	$-$17.523	&	5.434		&	51.621		\\
			12		&	19.165\,079	&	114.681	&	63.381		&	44.276		&	4.633		&	$-$18.697	&	5.973		&	55.159		\\
			13		&	19.164\,701	&	114.684	&	63.163		&	46.478		&	4.729		&	$-$19.539	&	6.417		&	57.776		\\
			14		&	19.164\,677	&	114.685	&	63.238		&	48.459		&	4.813		&	$-$20.301	&	6.807		&	60.137		\\
			15		&	19.164\,499	&	114.687	&	63.059		&	49.999		&	4.870		&	$-$20.854	&	7.137		&	61.905		\\
			16		&	19.164\,467	&	114.687	&	63.063		&	51.368		&	4.919		&	$-$21.340	&	7.428		&	63.459		\\
			17		&	19.164\,400	&	114.687	&	62.993		&	52.493		&	4.956		&	$-$21.717	&	7.679		&	64.704		\\
			18		&	19.164\,370	&	114.687	&	62.957		&	53.476		&	4.987		&	$-$22.036	&	7.901		&	65.768		\\
			19		&	19.164\,342	&	114.687	&	62.940		&	54.317		&	5.012		&	$-$22.297	&	8.097		&	66.658		\\
			20		&	19.164\,323	&	114.687	&	62.917		&	55.049		&	5.032		&	$-$22.515	&	8.271		&	67.417		\\
			21		&	19.164\,309	&	114.688	&	62.899		&	55.688		&	5.049		&	$-$22.698	&	8.427		&	68.066		\\
			22		&	19.164\,299	&		"		&	62.897		&	56.250		&	5.063		&	$-$22.851	&	8.567		&	68.623		\\
			23		&	19.164\,291	&		"		&	62.885		&	57.746		&	5.075		&	$-$22.981	&	8.692		&	69.105		\\
			24		&	19.164\,287	&		"		&	62.889		&	57.187		&	5.085		&	$-$23.091	&	8.806		&	69.525		\\
			25		&	19.164\,283	&		"		&	62.885		&	57.579		&	5.094		&	$-$23.185	&	8.909		&	69.891		\\
			26		&	19.164\,281	&		"		&	62.888		&	57.931		&	5.101		&	$-$23.266	&	9.003		&	70.214		\\
			27		&	19.164\,279	&		"		&	62.897		&	58.248		&	5.108		&	$-$23.335	&	9.088		&	70.500		\\
			28		&	19.164\,278	&		"		&	62.898		&	58.534		&	5.113		&	$-$23.396	&	9.167		&	70.754		\\
			29		&	19.164\,278	&		"		&	62.903		&	58.793		&	5.118		&	$-$23.448	&	9.239		&	70.982		\\
			30		&	19.164\,277	&		"		&		"		&	59.029		&	5.123		&	$-$23.495	&	9.305		&	71.186		\\
		\end{tabular}
	\end{ruledtabular}
\end{table}

We first discuss 2-hookium.  Choosing $M = 8$ yields the HF orbitals
\begin{subequations}
\begin{gather}
	\psi_1(x) = 0.989962 \,\phi_1(x) + 0.139577 \,\phi_3(x) - 0.021464 \,\phi_5(x) + 0.005740 \,\phi_7(x)	\\
	\psi_2(x) = 0.997679 \,\phi_2(x) + 0.067586 \,\phi_4(x) - 0.008026 \,\phi_6(x) + 0.001894 \,\phi_8(x)
\end{gather}
\end{subequations}
and Fig.~\ref{fig:hookium} reveals that the density and curvature are softened versions of those in 2-boxium.  As before, LDA1 interprets the density maxima as regions of strong correlation, predicting
\begin{equation}
	\EL = \int_{-\infty}^\infty \rho(x) \Ec^\text{LDA1}(\rs) \,dx = -42.2\ \mEh
\end{equation}
whereas gLDA1 finds that almost all of the correlation comes from a narrow region near the middle of the well and predicts
\begin{equation}
	\EgL = \int_{-\infty}^\infty \rho(x) \Ec^\text{gLDA1}(\rs,\eta) \,dx = -12.7\ \mEh
\end{equation}
As for 2-boxium, LDA1 and gLDA1 offer entirely different pictures of electron correlation but both perturbation theory ($\EP = -10.78\ \mEh$ and $\EPP = -12.66\ \mEh$) and near-exact calculations ($\ECI = -13.55\ \mEh$) agree that gLDA1 is closer to the truth.

We have also performed HF, LDA1, gLDA1, MP2, MP3 and FCI calculations on 3-, 4- and 5-hookium and the density and curvature for 5-hookium are shown on the right of Fig.~\ref{fig:hookium}.  As before, both functions oscillate more rapidly but with smaller amplitude than in 2-hookium and it is clear that, as the number of electrons becomes large, both functions will become increasingly uniform.\cite{Jellook12}

\begin{table}
	\caption{$\EHF$ and HOMO--LUMO gap (in $\Eh$) and $E_c$ (in $\mEh$) in $n$-boxium and $n$-hookium} \label{tab:apps}
	\begin{ruledtabular}
		\begin{tabular}{cccccccccc}
						&				\mc{4}{c}{$n$-boxium ($L=\pi$)}				&&				\mc{4}{c}{$n$-hookium ($k = 1$)}				\\
													\cline{2-5}														\cline{7-10}
						&	$n = 2$		&	$n = 3$		&	$n = 4$		&	$n = 5$		&&	$n = 2$		&	$n = 3$		&	$n = 4$		&	$n = 5$		\\
			\hline
			$\EHF$		&	3.48451	&	10.37969	&	22.42489	&	40.79205	&&	2.74367	&	6.63671	&	12.12335	&	19.16428	\\
			H-L gap	&	4.01		&	5.28		&	6.47		&	7.61		&&	1.75		&	1.72		&	1.69		&	1.67		\\
			$-\EL$		&	46.1		&	72.5		&	99.4		&	126.5		&&	42.2		&	65.9		&	90.1		&	114.7		\\
			$-\EgL$	&	11.0		&	26.3		&	44.0		&	63.0		&&	12.7		&	28.0		&	44.9		&	62.9		\\
			$-\EP$		&	8.3			&	23.1		&	41.8		&	62.8		&&	10.8		&	26.0		&	43.7		&	63.0		\\
			$-\EPP$	&	9.5			&	25.6		&	45.4		&	67.3		&&	12.7		&	30.0		&	49.8		&	71.1		\\
			$-\ECI$		&	9.8			&	26.2		&	46.1		&	68.0		&&	13.5		&	31.8		&	52.4		&	74.3		\\
		\end{tabular}
	\end{ruledtabular}
\end{table}

The convergence of the 5-hookium energies is shown in Table \ref{tab:convhook}.  As in 5-boxium, the LDA1 energies converge most rapidly, followed by the HF and gLDA1 energies, then the $O^4 V^2$, $O^3 V^3$ and $O^2 V^4$ components of the third-order energy, and finally the MP2 energy.  However, each of these energies converges significantly more slowly than its 5-boxium analog.  All of these observations are consistent with the theoretical predictions of Table \ref{tab:convergence}.  Because the $n$-hookiums are smaller-gap systems, MP2 and MP3 are less successful than for $n$-boxium, recovering roughly 85\% and 96\% of the correlation energy in 5-hookium.

Our best estimates of the CBS limit HF and correlation energies are summarized in the right half of Table \ref{tab:apps}.  As before, whereas LDA1 seriously overestimates the correlation energies, gLDA1 is within 15\% of the true correlation energy in all cases.  It is interesting to note that $|E_c$($n$-hookium)$|$ $>$ $|E_c$($n$-boxium)$|$ in all cases but that, whereas gLDA1 correctly predicts this trend, LDA1 reverses it.

\section{Concluding Remarks}
The traditional Local Density Approximation (LDA) is exact by construction for an infinite uniform electron gas with Seitz radius $\rs$.  However, it significantly overestimates the magnitudes of correlation energies in finite gases, such as those created when $n$ electrons are placed on the surface of a $\mD$-dimensional sphere.  This overestimation, which becomes even more pronounced in non-uniform gases, led us to seek generalizations of the LDA which are exact for both infinite and finite gases and, in the present work, we have proposed that the local hole curvature $\eta$ provides the necessary information to achieve this goal.  For present purposes, we have extracted $\eta$ from the HF wave function: this requires only the occupied HF orbitals.

By fitting accurately calculated correlation energies for systems of $n$ electrons on a ring, we have constructed the Generalized Local Density Approximation for one-dimensional systems and this has yielded a correlation kernel $\Ec(\rs,\eta)$ and a corresponding functional which we call GLDA1.  To this point, we have considered only gases in which $\eta \le 1$ and, consequently, the GLDA1 functional is not yet defined for gases with higher curvature.  However, if we assume that the the correlation kernel becomes flat, \textit{i.e.}~that $\Ec(\rs,\eta) = \Ec(\rs,1)$ when $\eta > 1$, we obtain an approximation to GLDA1 which we call gLDA1.

We have applied the traditional LDA1 functional and the curvature-corrected gLDA1 functional to electrons trapped in 1D boxes or in 1D harmonic wells and, by comparing the predicted correlation energies with those obtained from MP2, MP3 and Full CI calculations, we have discovered that gLDA1 is much more accurate than LDA1 in all cases.

We have also observed that gLDA1 tends to underestimate the magnitudes of correlation energies.  This suggests that the true GLDA1 kernel continues to rise, \textit{i.e.}~that $|\Ec(\rs,\eta)| > |\Ec(\rs,1)|$ but systematic examination of high-curvature ($\eta > 1$) gases is required to test this.  Such exploration is an important topic for future research and will allow the GLDA1 functional to be completely defined and tested.

Although we have presented relatively few calculations here, and much more investigation is warranted, these preliminary results suggest that ``curvature-corrected density functional theory (CC-DFT)'' may offer an efficient pathway to improvements over existing functionals.

\acknowledgments
P.F.L. and P.M.W.G. thank the NCI National Facility for generous grants of supercomputer time.  P.M.W.G. thanks the Australian Research Council (Grant Nos.~DP0984806, DP1094170 and DP120104740) for funding, P.F.L. thanks the Australian Research Council for a Discovery Early Career Researcher Award (Grant No.~DE130101441) and C.J.B. is grateful for an Australian Postgraduate Award.  P.F.L. and P.M.W.G. thank Neil Drummond and David Tew for helpful discussions and P.M.W.G. thanks the University of Bristol for sabbatical hospitality during the construction of this manuscript.

\newpage

\section*{Appendix A:  Calculation of matrix elements}
The matrix elements in Eq.~\eqref{eq:STV} are expressed as expectation values of operators over the HF wave function.  Therefore, because $\Phi^2$ and its reduced density matrices, \textit{e.g.}
\begin{equation} \label{eq:rho2}
	\rho_2(\theta_1,\theta_2) = \rho(\br)^2 \left( 1 - \left[ \frac{\sin n(\theta_1-\theta_2)/2}{n \sin(\theta_1-\theta_2)/2} \right]^2 \right)
\end{equation}
have \emph{finite} Fourier expansions, integrals of their products with Fourier expansions of operators reduce to \emph{finite} sums.

The Fourier expansions of bounded operators on a unit ring are straightforward, \textit{e.g.}
\begin{gather}
	r_{12}^2 = 2 - 2\cos(\theta_1-\theta_2)																		\\
	\nabla r_{12} \cdot \nabla r_{12} = 1 + \cos(\theta_1-\theta_2)												\\
	r_{12} = - \frac{4}{\pi} \sum_{a=-\infty}^\infty \frac{e^{i a(\theta_1-\theta_2)}}{4a^2-1}	\label{eq:rijexp}	\\
	\nabla r_{12} \cdot \nabla r_{13} =	\left[ \frac{4i}{\pi} \sum_{a=-\infty}^\infty \frac{a e^{i a(\theta_1-\theta_2)}}{4a^2-1} \right]
										\left[ \frac{4i}{\pi} \sum_{b=-\infty}^\infty \frac{b e^{i b(\theta_1-\theta_3)}}{4b^2-1} \right]	
\end{gather}
The expansions of unbounded operators, \textit{e.g.}
\begin{equation}
	r_{12}^{-1} = - \frac{2}{\pi} \sum_{a=-\infty}^\infty \left( \sum_{p=1}^{|a|} \frac{1}{2p-1} \right) e^{i a(\theta_1-\theta_2)}
\end{equation}
are delicate (they converge only in the Ces\`aro mean\cite{NISTbook}) but this is sufficient for our purposes because we require only a few of the low-order Fourier coefficients.  The expansions of ``cyclic'' operators (\textit{e.g.}~$r_{12}r_{23}r_{31}$) are not simple products and must be derived separately.

Thus, for example, to find the $\langle \Phi | r_{12} r_{13} | \Phi \rangle$ integral in 3-ringium, the Fourier expansion
\begin{equation}
	\Phi^2 = \frac{[2-2\cos(\theta_1-\theta_2)] \, [2-2\cos(\theta_1-\theta_3)] \, [2-2\cos(\theta_2-\theta_3)]} {3!(2\pi)^3}
\end{equation}
is combined with Eq.~\eqref{eq:rijexp} to yield
\begin{align}
	\langle \Phi | r_{12}r_{13} | \Phi \rangle	& = \frac{16}{\pi^2} \sum_{a=-2}^2 \sum_{b=-2}^2 \iiint
												\frac{e^{i a(\theta_1-\theta_2)}}{4a^2-1}
												\frac{e^{i b(\theta_1-\theta_3)}}{4b^2-1} \Phi^2 d\theta_1 d\theta_2 d\theta_3	\notag	\\
											& = \frac{16384}{675\pi^2}
\end{align}

\section*{Appendix B:  Extrapolation of perturbation energies}
It is common these days to estimate the CBS limit of post-HF correlation energies by extrapolation.\cite{Helgaker97}  Pioneering work by Schwartz,\cite{Schwartz62} Hill\cite{Hill85} and Kutzelnigg and Morgan\cite{Kutzelnigg92} showed that, for atoms in 3D, the second-order energy contributions from basis functions with angular momentum $\ell$ converge asymptotically as $(\ell+1/2)^{-4}$.

While generating the data in Section \ref{sec:valid}, we found that the MP2 and MP3 energies converge so slowly (Tables \ref{tab:convbox} and \ref{tab:convhook}) that the CBS limit is not reached (within our 0.1 $\mEh$ target accuracy), even with our largest ($M = 30$) basis set.  This is particularly noticeable for $n$-hookium.  We therefore needed to develop and apply appropriate extrapolation procedures.

To this end, we analyzed the convergence of the second-order energy
\begin{equation}	\label{eq:PT2}
	E^{(2)} = \sum_{r=3}^\infty \sum_{s=r+1}^\infty \frac{\langle 12||rs \rangle^2}{\epsilon_1+\epsilon_2-\epsilon_r-\epsilon_s}
\end{equation}
obtained from the non-interacting orbitals and orbital energies in 2-boxium and 2-hookium.  In $n$-hookium, the double-bar integral is
\begin{equation}
	\langle 12 || rs \rangle = \frac{(-1)^{(r-s+1)/2}\sqrt{2}}{\pi} \ \frac{\G((r+s-2)/2)}{\sqrt{\G(r)\G(s)}}
\end{equation}
if $r+s$ is odd but it vanishes if $r+s$ is even.  The orbital energies are given by $\epsilon_k = k - 1/2$.  By substituting these expressions into \eqref{eq:PT2} and making use of Stirling's approximation,\cite{NISTbook} one can show that the error introduced by truncating the basis after $M$ functions is
\begin{align}
	\Delta E^{(2)}_M	& = \sum_{r=3}^M \sum_{s=r+1}^M \frac{\langle 12||rs \rangle^2}{\epsilon_1+\epsilon_2-\epsilon_r-\epsilon_s} - E^{(2)}	\notag	\\
						& \sim \frac{1}{3(\pi M)^{3/2}} + O(M^{-2})
\end{align}
The closed-form expression for the $\langle 12 || rs \rangle$ integral in $n$-boxium is cumbersome but a similar analysis reveals that the analogous truncation error is $O(M^{-3})$.  The truncation errors in the third-order energy can be found in the same way and all of our results are summarized in Table \ref{tab:convergence}.

The MP2, MP3 and FCI energies obtained with our largest basis sets conform to these analytical predictions and allowed us to extrapolate reliably to the CBS energies given in Table \ref{tab:apps}.  The good agreement between our extrapolated FCI energies and QMC energies further increases our confidence in these results.

\end{document}